\documentclass{emulateapj}

\usepackage{natbib}
\def\farcs{\hbox{$.\!\!^{\prime\prime}$}}
\def\degr{\hbox{$^\circ$}}\def\arcmin{\hbox{$^\prime$}}
\def\arcsec{\hbox{$^{\prime\prime}$}}
\newcommand{\oi}{[\ion{O}{1}]}
\newcommand{\cii}{[\ion{C}{2}]}
\newcommand{\neii}{[\ion{Ne}{2}]}
\newcommand{\ho}{H$_2$O}
\newcommand{\kms}{km\,s$^{-1}$}

\newcommand{\um}{$\rm \mu$m}
\newcommand{\ergs}{erg\,s$^{-1}$}

\newcommand{\wm}{W\,m$^{-2}$}
\newcommand{\cmc}{cm$^{-3}$}

\newcommand{\msol}{$M_{\odot}$} 
\newcommand{\lsol}{$L_{\odot}$} 
 
\newcommand{\be}{\begin{equation}} 
\newcommand{\ee}{\end{equation}} 

\begin{document}

\title{Probing the gaseous disk of T Tau N with CN 5--4 lines}

\author{L. Podio\altaffilmark{1,2}, I. Kamp\altaffilmark{3}, C. Codella\altaffilmark{1}, B. Nisini\altaffilmark{4},
G. Aresu\altaffilmark{5}, S. Brittain\altaffilmark{6}, S. Cabrit\altaffilmark{7,2}, C. Dougados\altaffilmark{8,2},
C. Grady\altaffilmark{9,10}, R. Meijerink\altaffilmark{3,11}, G. Sandell\altaffilmark{12}, M. Spaans\altaffilmark{3},  
W.-F. Thi\altaffilmark{2},  G. J. White\altaffilmark{13,14},  P. Woitke\altaffilmark{15} 
}

\altaffiltext{1}{INAF-Osservatorio Astrofisico di Arcetri, Largo E. Fermi 5, 50125, Florence, Italy} 
\altaffiltext{2}{UJF-Grenoble 1/CNRS-INSU, Institut de Plan\'etologie et d'Astrophysique de Grenoble (IPAG) UMR 5274, Grenoble, F-38041, France}
\altaffiltext{3}{Kapteyn Astronomical Institute, University of Groningen, Landleven 12, 9747 AD Groningen, The Netherlands} 
\altaffiltext{4}{INAF-Osservatorio Astronomico di Roma, via di Frascati 33, 00040, Monte Porzio Catone, Italy}  
\altaffiltext{5}{INAF-Osservatorio Astronomico di Cagliari, Via della Scienza 5, 09047 Selargius, Italy}  
\altaffiltext{6}{Department of Physics \& Astronomy, 118 Kinard Laboratory, Clemson University, Clemson, SC 29634, USA}
\altaffiltext{7}{LERMA, Observatoire de Paris, UMR 8112 CNRS/INSU, 61 Av. de l'Observatoire, 75014, Paris, France}
\altaffiltext{8}{LFCA, UMI 3386, CNRS and Dept. de Astronomia, Universidad de Chile, Santiago, Chile}
\altaffiltext{9}{Eureka Scientific, 2452 Delmer, Suite 100, Oakland, CA 96002, USA}
\altaffiltext{10}{Exoplanets \& Stellar Astrophysics Laboratory, NASA Goddard Space Flight Center, Code 667, Greenbelt, MD 20771, USA} 
\altaffiltext{11}{Leiden Observatory, Leiden University, P.O. Box, NL-2300 RA Leiden, The Netherlands}
\altaffiltext{12}{SOFIA-USRA, NASA Ames Research Center, MS 232-12, Building N232, Rm. 146, P.O. Box 1, Moffett Field, CA 94035-0001, USA}
\altaffiltext{13}{Department of Physical Sciences, The Open University, Milton Keynes MK7~6AA, UK}  
\altaffiltext{14}{RALSpace, The Rutherford Appleton Laboratory, Chilton, Didcot, OX11 0QX, UK}  
\altaffiltext{15}{SUPA, School of Physics and Astronomy, University of St. Andrews, KY16 9SS, UK}  

\begin{abstract}
We present spectrally resolved {\it Herschel}/HIFI observations of the young multiple system T~Tau in atomic and molecular lines. 
While CO, \ho, \cii, and SO lines trace the envelope and the outflowing gas up to velocities of 33 \kms\ with respect to systemic, the CN 5--4 hyperfine structure lines at 566.7, 566.9 GHz show a narrow double-peaked profile centered at systemic velocity, consistent with an origin in the outer region of the compact disk of T Tau N.
Disk modeling of the T Tau N disk with the thermo-chemical code ProDiMo produces CN line fluxes and profiles consistent with the observed ones and constrain the size of the gaseous disk ($R_{\rm out}=110^{+10}_{-20}$~AU) and its inclination ($i=25$\degr$\pm5$\degr). 
The model indicates that the CN lines originate in a disk upper layer at 40--110~AU from the star, which is irradiated by the stellar UV field and heated up to temperatures of $50-700$~K. 
With respect to previously observed CN 2--1 millimeter lines, the CN 5--4 lines appear to be less affected by envelope emission, due to their larger critical density and excitation temperature. 
Hence, high-J CN lines are a unique confusion-free tracer of embedded disks, such as the disk of T Tau N.
\end{abstract}


\section{Introduction}

\begin{table*}
\caption{\label{tab:lines} Observed lines}
\begin{centering}
\begin{tabular}{ccccccccccc}
\hline\hline
Transition & $\nu_{\rm 0}$$^a$ & $E_{\rm up}$ & HPBW & $T_{\rm peak}$ & $V_{\rm peak}$ & $\Delta V$ & $\int T_{\rm mb} dV$ & $F_{\rm obs}$ & $F_{\rm mod}$\\
                &                      GHz &                 K & \arcsec &          mK          &      \kms       & \kms &             K \kms &        \wm &   \wm    \\
\hline
                   CN 5--4 9/2--7/2 &  566.731 &   82 & 37 &  21/25 $\pm$ 6$^b$ & 6.7/9.1 $\pm$ 0.5$^b$ &  7 &  0.08 $\pm$  0.01 & 5.4 $\pm$ 0.6 10$^{-19}$ & 5.7 10$^{-19}$\\ 
                  CN 5--4 11/2--9/2 &  566.947 &   82 & 37 &  25/28 $\pm$ 6$^b$ & 6.3/8.4 $\pm$ 0.5$^b$ &  7 &  0.12 $\pm$  0.01 & 8.1 $\pm$ 0.6 10$^{-19}$ & 6.7 10$^{-19}$\\ 
        o-\ho\ 1$_{10}$--1$_{01}$ &  556.936 &   61 & 38 &                                   682 $\pm$ 6 &                          8.6 $\pm$ 0.5 & 52 &  5.40 $\pm$  0.02 & 3.70 $\pm$ 0.02 10$^{-17}$ & 3.5 10$^{-19}$\\ 
        p-\ho\ 1$_{11}$--0$_{00}$ & 1113.343 &   53 & 19 &                                   873 $\pm$ 8 &                          8.6 $\pm$ 0.3 & 52 &  9.85 $\pm$  0.02 & 1.344 $\pm$ 0.003 10$^{-16}$ & 2.2 10$^{-18}$\\ 
        o-\ho\ 3$_{12}$--2$_{21}$ & 1153.127 &  249 & 18 &                                     900  $\pm$ 75  &                          6.2 $\pm$ 0.3 & 21 &  7.2  $\pm$  0.1  & 1.02 $\pm$ 0.02 10$^{-16}$ & 6.9 10$^{-19}$\\ 
                          CO 10--9 & 1151.985 &  304 & 18 &                       13210 $\pm$ 75 &  7.7 $\pm$ 0.3 & 38 & 82.9 $\pm$  0.2 & 1.171 $\pm$ 0.002 10$^{-15}$ & 1.4 10$^{-17}$\\ 
                   $^{13}$CO 10--9 & 1101.350 &  291 & 19 &                    1248 $\pm$ 8 & 7.7 $\pm$ 0.3 & 20 &  4.92 $\pm$  0.01 & 6.64 $\pm$ 0.02 10$^{-17}$ & 3.1 10$^{-18}$\\ 
          SO 12$_{13}$--11$_{12}$ &  558.087 &  194 & 38 &                  19 $\pm$ 6 &  7.2 $\pm$ 0.5 &  9 &  0.08 $\pm$  0.01 & 5.3 $\pm$ 0.7 10$^{-19}$ & 7.5 10$^{-20}$\\ 
\cii\ $^2$P$_{3/2}$--$^2$P$_{1/2}$ & 1900.537 &   91 & 11 &                    1120 $\pm$ 83 &  7.1 $\pm$ 0.2 & 30 &  8.4 $\pm$  0.1 & 1.95 $\pm$ 0.03 10$^{-16}$ & 4.8 10$^{-17}$\\ 
%
\hline
\end{tabular}\\
$^a$Frequencies are from the Jet Propulsion Laboratory database (JPL, \citealt{pickett98}) and from the Cologne Database of Molecular Spectroscopy (CDMS, \citealt{muller01}) for CN\\
$^b$For the double-peaked CN lines the intensity and velocity of both peaks is reported\\
\end{centering}
\end{table*}

  \begin{figure}
     \centering
     \includegraphics[width=8.8cm]{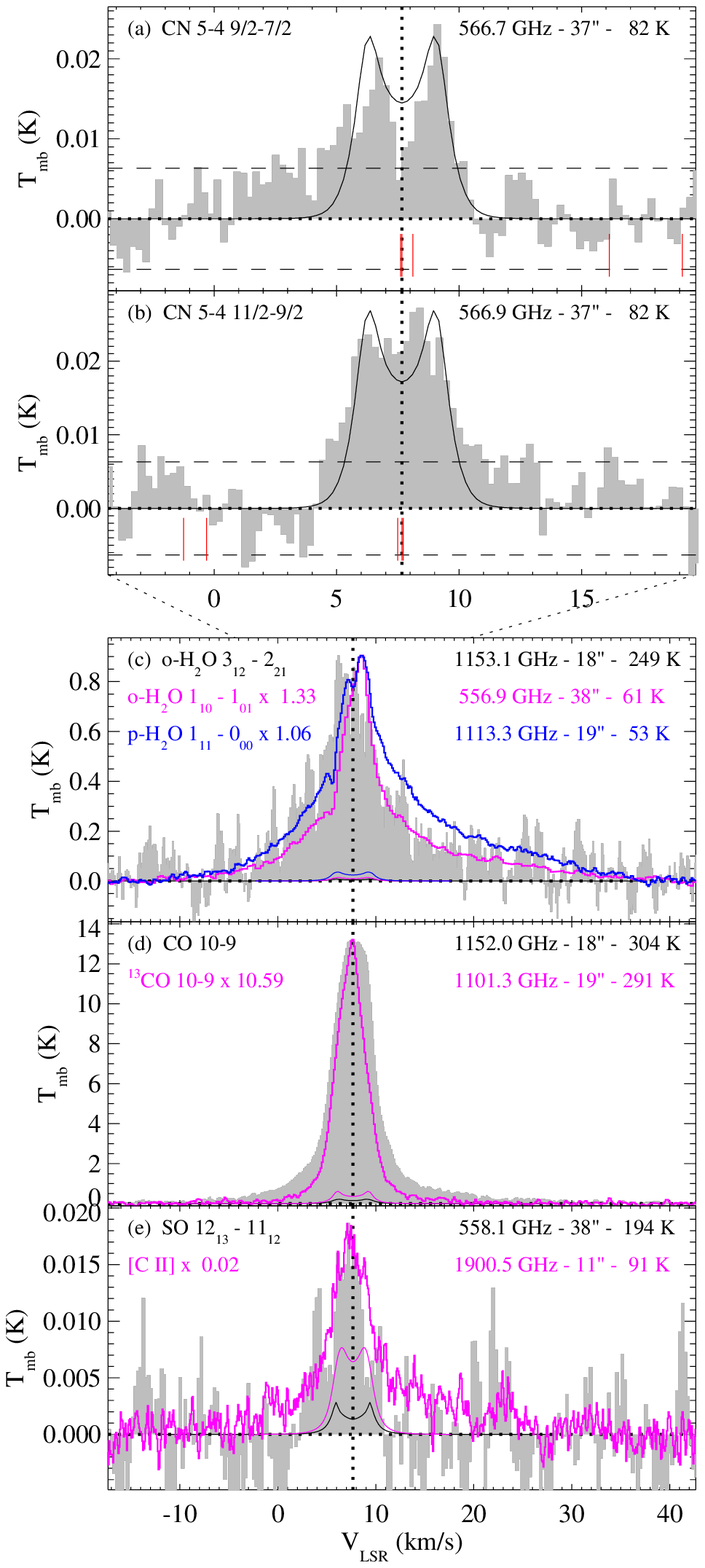}
   \caption{{\it Herschel}/HIFI spectra of CN 5--4 9/2--7/2 ({\it panel a}), CN 5--4 11/2--9/2 ({\it panel b}), o-\ho~3$_{12}$-2$_{21}$, o-\ho~1$_{10}$-1$_{01}$, p-\ho~1$_{11}$-0$_{00}$ ({\it panel c}), CO 10--9, $^{13}$CO 10--9 ({\it panel d}), SO 12$_{13}$--11$_{12}$, \cii~$^2$P$_{3/2}$-$^2$P$_{1/2}$ ({\it panel e}). Horizontal and vertical dotted lines indicate the baseline and systemic velocity ($V_{\rm sys}=+7.7$~\kms). 
The line frequency,  HPBW, and upper level energy are labelled. 
For CN 5--4 the rms noise is indicated by the dashed horizontal line and the positions of the hyperfine structure lines by the red vertical lines.
The line profiles predicted by the ProDiMo disk model of T Tau N are overplotted 
in black, magenta, and blue (corresponding observed profiles are in grey, magenta, and blue).
}
   \label{fig:hifi_lines}
    \end{figure}

The study of  the physical and chemical structure of protoplanetary disks is crucial to comprehend the formation of planetary systems. 
According to models disks have a stratified structure, hence different molecular species probe the physical and chemical conditions of different layers from the warm irradiated surface down to the cold midplane. 
Disks around non-embedded T Tauri and Herbig stars have been imaged in CO \citep[e.g., ][]{dutrey98,qi04,pietu07}, and detected in a number of other millimeter (mm) and sub-mm molecular lines \citep[e.g., ][]{dutrey97,thi04}.
On the contrary, the study of disks around more embedded systems is obstacled by the associated envelopes and outflows which also emit strongly in the same molecular lines, hiding the fainter disk emission.  

Recent studies show that CN is a good disk tracer with 88\% of the T Tauri disks detected in CO 2--1 which are also observed in the CN 2--1  lines \citep{oberg10,oberg11,chapillon12}.
Moreover, \citet{guilloteau13} performed an IRAM 30-m survey of T Tauri and Herbig Ae systems located mainly in the Taurus-Auriga region and show that the CN 2--1 lines are less affected by the emission from the surrounding molecular cloud than the lines from CO isotopologues.
However, CN 2--1 lines are still dominated by envelope/outflow emission in the case of actively accreting/ejecting sources, such as T~Tau.

T~Tau is a multiple system driving at least two bipolar jets detected in optical, near-infrared forbidden lines \citep{bohm94,eisloffel98,solf99,herbst07}. 
The system consists of the northern component T Tau N ($M_{*}=2.1$~\msol) and of the binary system T Tau Sa+Sb located 0\farcs7 to the south ($M_{*}=2.1,0.8$~\msol, separation Sa--Sb=0$\farcs$13) \citep{kohler08}.
T Tau N is optically visible and is surrounded by an almost face-on, intermediate mass disk ($i < 30$\degr,  $M_{\rm disk} \sim 0.01$ \msol) and an optically thin envelope \citep{beckwith90,hogerheijde97,akeson98,ratzka09,guilloteau11}. 
Both southern components, instead, are strongly obscured ($A_{\rm V}\simeq15,30$ mag towards T Tau Sb and Sa, respectively), possibly due to a circumbinary envelope and, for T Tau Sa, to its almost edge-on disk  ($i\sim72$\degr, \citealt{duchene05,ratzka09}).
Sa and Sb are not detected at mm wavelengths suggesting that their disks are small and of low-mass ($M_{\rm disk}\sim10^{-5}-10^{-3}$~\msol,  $R_{\rm out}\sim5$~AU, \citealt{hogerheijde97,ratzka09}). 
While the dusty disk of T Tau N has been mapped in the mm, there are only very tentative observations of its gaseous disk, e.g. in the H$_2$ near-infrared lines \citep{gustafsson08}.  
Emission in the \oi~63~\um, \cii~157~\um, and \neii~12.81~\um\ lines is extended and/or velocity-shifted, hence likely associated with outflowing gas, while molecular lines, such as high-J CO, OH, and \ho\ lines, are spectrally and spatially unresolved  \citep{spinoglio00,vanboekel09,podio12}.
Observations of low-J CO lines \citep{edwards82,schuster93} as well as of less abundant molecular species, such as $^{13}$CO, C$^{17}$O, o-H$_2$CO, SO, and CN 2--1 are also dominated by envelope/outflow emission, preventing us to trace the gas in the disk \citep{guilloteau13}.

In this letter we present {\it Herschel}/HIFI observations of the T Tau system in atomic and molecular lines with higher excitation energies than probed by \citet{guilloteau13} in the mm range. 
While CO, $^{13}$CO, \ho, \cii, and SO lines are still strongly dominated by emission from the surrounding envelope and outflows, the CN 5--4 lines allow digging out the faint emission from the disk of T Tau N.

\section{Observations and data reduction}
\label{sect:obs}

We observed T Tau ($\alpha_{\rm J2000}=04^{\rm h}$ 21$^{\rm m}$ 59$\fs$4, $\delta_{\rm J2000}=+19\degr$ 32$\arcmin$ 06$\farcs$4) with the Heterodyne Instrument for the Far Infrared (HIFI, \citealt{degraauw10}) on board the {\it Herschel}  Space Observatory\footnote{{\it Herschel} is an ESA space observatory with science instruments provided by European-led Principal Investigator consortia and with important participation from NASA.} \citep{pilbratt10}. 
The observations target the two fundamental water lines, o-\ho~1$_{10}$--1$_{01}$ and p-\ho~1$_{11}$--0$_{00}$ in the HIFI bands 1 and 4, the $^{12}$CO (hereafter CO) 10--9 line in band 5, and the \cii~$^2$P$_{3/2}$--$^2$P$_{1/2}$ line in band 7 (OBSID: 1342249419, 1342250207, 1342249598, 1342249647).
They were acquired with a single on-source pointing and in dual beam switch mode with fast chopping 3\arcmin\ either side of the target. 
The achieved rms noise in bands $1,4,5,7$ is of $6,8,75,83$ mK, respectively.
The Wide Band Spectrometer (WBS) and the High Resolution Spectrometer (HRS) were used in parallel, providing a spectral resolution of 1.10 and 0.25~MHz, respectively.
The Half Power Beam Width (HPBW) ranges from $\sim11$\arcsec\ to $\sim38$\arcsec, depending on frequency. 
Hence all the three sources of the T Tau system (N, Sa, and Sb) are covered by the observations.
The selected spectral settings also cover with the WBS the CN 5--4 J=9/2--7/2 and J=11/2--9/2, SO 12$_{13}$--11$_{12}$, $^{13}$CO 10--9, and o-\ho\ 3$_{12}$--2$_{21}$ lines. 

The HIFI data were reduced using HIPE 10\footnote{HIPE is a joint development by the {\it Herschel} Science Ground Segment Consortium, consisting of ESA, the NASA {\it Herschel} Science Center, and the HIFI, PACS and SPIRE consortia.}.
Fits files from level 2 were then created and transformed into GILDAS\footnote{http://www.iram.fr/IRAMFR/GILDAS} format for data analysis. The spectra were baseline subtracted and averaged over the horizonthal and vertical polarization to increase the signal-to-noise ratio.
Antenna temperatures, $T_{\rm a}$, were converted to mean beam temperature, $T_{\rm mb}$, using mean beam efficiency by \citet{roelfsema12}. 

The properties of the detected lines (transition, frequency, $\nu_{\rm 0}$, upper level energy, $E_{\rm up}$), the HPBW, and the measured parameters (peak temperature and velocity, $T_{\rm peak}$ and $V_{\rm peak}$, the full line width at 1$\sigma$ rms noise, $\Delta V$, integrated intensity, $\int T_{\rm mb}dV$, and line fluxes, $F_{\rm obs}$\footnote{$F_{\rm obs}=\frac{2K_{\rm b}\nu^3}{c^3}\times\int T_{\rm mb}dV\times\pi\left(\frac{HPBW}{2\sqrt{\ln2}}\right)^2$}) are summarized in Table \ref{tab:lines}.

\section{Results from observations}
\label{sect:results}

The detected lines in Figure \ref{fig:hifi_lines} show a variety of profiles indicating that they probe different gas components in the complex T Tau system. 

The CO and $^{13}$CO 10--9 lines show a single-peaked profile, likely dominated by cloud emission, and red-shifted and blue-shifted wings with velocities up to $-8$ and $+30$ \kms\ probing outflowing gas.  
Assuming an average local ISM carbon isotope ratio of $68\pm15$ \citep{milam05}, the $^{12}$CO/$^{13}$CO line ratio indicates that the CO 10--9 emission is optically thick, its optical depth varying between $6.7\pm1.5$ at the peak velocity down to $2.4\pm0.8$ for gas velocities of $\sim5-6$~\kms\ with respect to $V_{\rm peak}$. 
Since the $^{13}$CO 10--9 line is optically thin, it can be used to estimate the systemic velocity. 
The line peak indicates $V_{\rm sys}=+7.7\pm0.3$ \kms, consistent with previous estimates from $^{13}$CO and C$^{18}$O 1--0, 2--1 lines \citep{edwards82,schuster93,guilloteau13}.  

The \cii\ $^2$P$_{3/2}$--$^2$P$_{1/2}$ and SO 12$_{13}$--11$_{12}$ lines peak at slightly blue-shifted velocity with respect to systemic ($V_{\rm peak}=+7.1\pm0.2,+7.2\pm0.5$ \kms).
As previously suggested by \citet{guilloteau13} SO lines trace the envelope with no evidence of high-velocity wings.
The \cii\ line instead shows high-velocity blue-shifted and red-shifted wings associated with outflowing gas.

We also detect three water lines: the o-\ho~1$_{10}$--1$_{01}$, o-\ho~3$_{12}$--2$_{21}$, and p-\ho~1$_{11}$--0$_{00}$ lines. The two fundamental water lines  ($E_{\rm up}\sim61,53$~K) show an absorption feature at the systemic velocity, due to the cloud, and a second absorption at $\sim+5.6$ \kms, whose origin is unclear.
As already observed in previous works \citep[e.g., ][]{kristensen12}, water lines appear to be very sensitive to high-velocity emission showing wings extending up to velocities of $-12$ \kms\ and $+40$ \kms. 

In contrast with previous observations of CO 6--5, 3--2, 2--1, and 1--0 lines \citep{edwards82,schuster93}, the observed CO 10--9, \ho, and \cii\ lines show a red-shifted wing which is brighter and extends to higher velocities than the blue one.
The lines observed with HIFI have higher upper level energies and/or critical densities than previously observed low-J CO lines ($E_{\rm up}\le116$~K), thus suggesting that the red-shifted outflowing gas is denser and more excited than the blue-shifted gas.

Our observations also cover the CN N=5--4 hyperfine structure lines\footnote{Due to the coupling of the rigid-body angular momentum, N, with the electronic spin, S, and with the nuclear spin, I, angular momenta, the CN N=5--4 line is split into 19 hyperfine components characterised by the corresponding quantum numbers J=N+S, and F=J+I.} \citep{muller01}.
We detect the three brightest J=9/2--7/2 and J=11/2--9/2 lines at 566.7~GHz and 566.9~GHz,  while the fainter J=9/2--9/2 components fall outside the observed range. 
The separation in velocity between the three detected lines is between 0.05--0.45 \kms, hence the lines are not resolved at the available resolution ($0.5$ \kms) and the observed profile is sensitive {\it only} to the kinematics of the emitting gas.
The CN 5--4 J=9/2--7/2 lines show a narrow double-peaked profile centered at the systemic velocity with a total line width of $\sim7$~\kms, FWHM of $4.0\pm0.5$~\kms, and a peak separation $\Delta V_{\rm sep}=2.4\pm0.5$~\kms.
The profile of the J=11/2--9/2 blended components is also narrow and symmetric around the systemic velocity, even though the double-peak is not as clear as in the  J=9/2--7/2 lines (see Table \ref{tab:lines}). 
The blue-/red- shifted wing shown by the J=9/2--7/2 and the J=11/2--9/2 components, respectively, is clearly due to the noise as it is below the noise level and is not detected in both components. 
Hence, 
the CN 5--4 profiles are consistent with emission from the outer region of the disk of T Tau N.
The double-peak detected in the CN 5--4 lines is not seen in the 2--1 lines observed with the IRAM-30m by \citet{guilloteau13}. This is due to the higher excitation energy and critical density  of the CN 5--4 lines ($E_{\rm up}\sim82$ K, n$_{\rm cr}\sim2~10^{8}$ \cmc), which makes them less affected by cloud emission, hence more effective to probe the disk, than CN 2--1  ($E_{\rm up}\sim16$ K, n$_{\rm cr}\sim2~10^{7}$ \cmc). 
Emission from the disks of T Tau Sa and T Tau Sb is expected to be negligible with respect to the emission from the disk of T Tau N, as those disks are not detected in the mm continuum and are one to two orders of magnitude less massive and very small \citep{akeson98,ratzka09,guilloteau11}. 

Assuming Keplerian rotation, a stellar mass of 2.1~\msol\ and an inclination of $i\simeq20-30$\degr\ \citep{ratzka09}, the peak separation of the CN lines indicates an outer disk radius $R_{\rm out}$(CN)$< 160-350$~AU. 
This upper limit is in agreement with the size of the disk as estimated from continuum maps at 1.3, 2.7 mm ($R_{\rm out}$(dust)$=67\pm20$~AU, \citealt{guilloteau11}) and from the H$_2$ ring-like structure observed by \citet{gustafsson08} ($R_{\rm out}$(H$_2$)$=85-100$~AU). 

\section{Modeling CN lines from the disk of T Tau N}
\label{sect:modeling}

\begin{table}
\begin{center}
\caption{\label{tab:model_par} T~Tau~N disk model: star and disk parameters}
    \begin{tabular}{lll}
\hline\hline
Effective temperature & $T_{\rm eff}$ (K)        & 5250  \\
Stellar mass                & $M_{*}$ (\msol)       & 2.1    \\       
Stellar luminosity       & $L_{*}$ (\lsol)          & 7.3       \\    
UV excess                   & $f_{\rm UV}$              & 0.1    \\
UV power law index    & $p_{\rm UV}$             & 0.2   \\             
X-rays luminosity       & $L_{\rm X}$ (\ergs)    & 2~10$^{31}$ \\ 
Disk inclination          & $i$ (\degr)              & 25 \\
Disk inner radius        & $R_{\rm in}$ (AU)        & 0.1   \\
Disk outer radius        & $R_{\rm out}$ (AU)       & 110    \\
Disk dust mass           & $M_{\rm dust}$ (\msol) & 1.3~10$^{-4}$ \\
Dust-to-gas ratio       &  dust-to-gas             & 0.01 \\ 
Dust material mass density & $\rho_{\rm dust}$ (g~\cmc) & 2.5 \\
Minimum grain size    & $a_{\rm min}$ (\um)    & 0.005 \\
Maximum grain size   & $a_{\rm max}$ (\um)    & 1000  \\
Dust size distribution index & $q$              & 3.5 \\
Surface density $\Sigma\approx r^{-\epsilon}$ & $\epsilon$ & 1 \\
Scale height at $R_{\rm in}$ & $H_{\rm 0}$ (AU)         & 0.0032 \\
Flaring index $H(r)=H_{\rm 0}\left(\frac{r}{R_{\rm in}}\right)^{\beta}$  & $\beta$ & 1.25 \\ 
Settling $H(r,a)=H(r)\frac{a}{a_{\rm set}}^{-s_{\rm set}/2}$ & $s_{\rm set}$  & 0.5 \\ %
Minimum grain size for settling & $a_{\rm set}$ (\um)  & 0.25 \\
Fraction of PAHs w.r.t. ISM & $f_{\rm PAH}$      & 0.01 \\
\hline
    \end{tabular}
\end{center}
\end{table}


   \begin{figure}
     \centering
     \includegraphics[width=8.8cm]{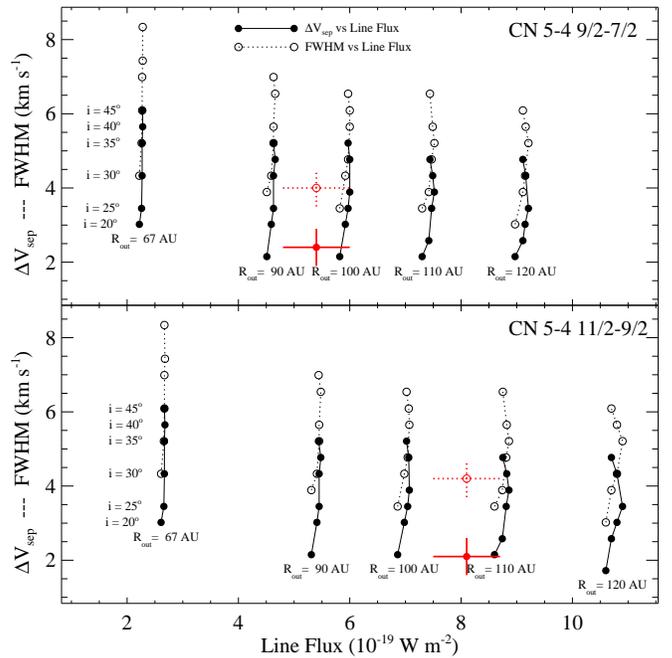}
   \caption{Observed line properties (in red) are compared with model predictions (in black) for different values of the disk outer radius, $R_{\rm out}$ (AU), and inclination, $i$ (\degr), for the CN 5--4 J=9/2--7/2 and J=11/2--9/2 lines (top and bottom panel, respectively). The peak separation versus the line flux is indicated by filled points and solid lines connecting the model predictions for the same  $R_{\rm out}$. The FWHM versus the line flux is overplotted with empty circles and dotted lines.}
   \label{fig:grid}
    \end{figure}

   \begin{figure}
     \centering
     \includegraphics[width=8.8cm]{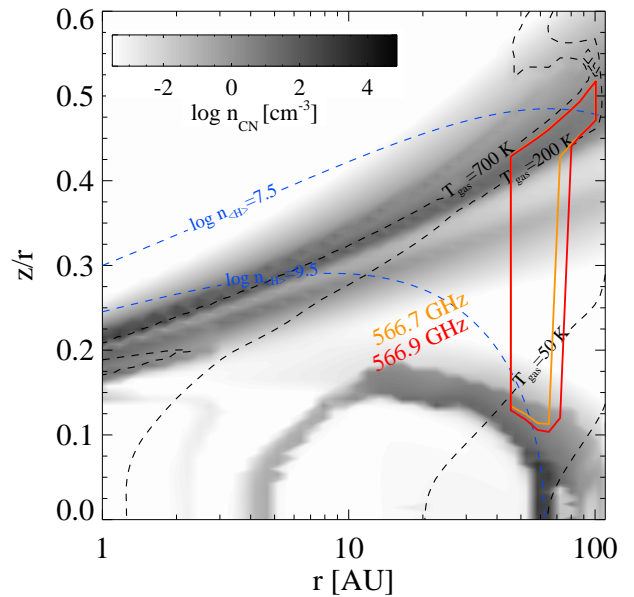}
   \caption{Disk region from which 50\% of the CN 5--4 J=9/2--7/2 at 566.7 GHz (in orange), and the CN 5--4 J=11/2--9/2 at 566.9 GHz (in red) line emission arises in the disk model of T Tau N.  Note that the emitting regions of the two CN lines overlap almost completely. The grey colour indicates the CN density, $n_{\rm CN}$ (\cmc), the dotted black and blue curves the gas temperature and total hydrogen number density $n_{\rm \langle H\rangle}=n_{\rm H}+2 n_{\rm H_2}$.}
   \label{fig:CN_model}
    \end{figure}


In order to test if the CN lines originate in the disk of T Tau N we use a parametrized disk model calculated with the thermo-chemical disk modeling code ProDiMo \citep{woitke09,kamp10,aresu11}. 

We adopt stellar and disk parameters as inferred from previous studies, which well reproduce the source spectral energy distribution (SED) \citep[e.g., ][]{ratzka09}.
We use stellar spectral type K0 ($T_{\rm eff}\simeq5250$~K), stellar luminosity $\simeq7.3$~\lsol\ and stellar mass $M_{\star}\simeq2.1$~\msol\ as determined by \citet{white01}. 
The UV spectrum \citep{calvet04} is reproduced by an UV excess $f_{\rm UV}=L(910-2500~\AA)/L_\ast=0.1$ and a power law slope $L_{\rm \lambda}\approx\lambda^{0.2}$. 
The effect of X-ray radiation from the stellar corona of T Tau N ($L_{\rm X}=2~10^{31}$~\ergs, \citealt{gudel07a}) is taken into account following \citet{aresu11} and \citet{meijerink12}. 
Following \citet{ratzka09} we assume a disk inner radius $R_{\rm in}=0.1$~AU, a dust mixture of astronomical silicates \citep{draine84} and amorphous carbon \citep{zubko96}  with relative abundances of 62.5\% and 37.5\%, and a grain size distribution $n(a)\approx a^{-q}$ with $q=3.5$, where $n(a)$ is the number of dust particles with radius $a$, and the minimum/maximum grain size are $a_{\rm min}=0.005$~\um\ and $a_{\rm max}=1$~mm. 
This implies a dust mass of $1.3~10^{-4}$ \msol\ to reproduce the observed mm emission \citep{hogerheijde97,akeson98,guilloteau11}.
Using the average ISM dust-to-gas ratio of $0.01$, the gas mass is 0.013~\msol.
The dust-to-gas ratio can be different from the ISM value in evolved disks \citep[e.g., ][]{thi10,gorti11}.
However, as the CN lines are optically thick in the disk, their flux is independent on the gas mass, hence on the assumed dust-to-gas ratio.
The SED is well reproduced by assuming a disk surface density $\Sigma\approx r^{-1}$ and scale height $H=0.0032~{\rm AU}~\left(r/R_{\rm in}\right)^{1.25}$ \citep{ratzka09}. 
To obtain a dust distribution similar to the two-layers model adopted by \citet{ratzka09}, we assume that grains larger than $a_{\rm set}=0.25$~\um\ have settled 
to a smaller scale height than the gas, $H(r,a)=H(r)\left(a/a_{\rm set}\right)^{-s_{\rm set/2}}$, with $s_{\rm set}=0.5$.
As PAH emission is generally not detected in TTSs \citep[e.g., ][]{furlan06}, the PAH fraction, $f_{\rm PAH}$, is set to $0.01$ with respect to the ISM abundance of $10^{-6.52}$ PAH particles/H-nucleus. Lower values, $f_{\rm PAH}=10^{-3}-10^{-4}$, do not affect the CN emission.
The adopted stellar and disk parameters are summarized in Table \ref{tab:model_par}.

As the CN line fluxes and profiles are very sensitive to the assumed disk outer radius,  $R_{\rm out}$, and inclination, $i$, we proceed through two steps.
First, we run a grid of models at low resolution (50$\times$50 grid points) adopting $R_{\rm out}$ and $i$ values which cover the range of estimates from previous studies ($R_{\rm out}=67,90,100,110,120$~AU, $i=20,25,30,35,40,45$ AU, \citealt{stapelfeldt98,gustafsson08,ratzka09,guilloteau11,harris12}).
Then, we run a high resolution model (100$\times$100 grid points) for $R_{\rm out}$ and $i$ which best fit the observations,
to better resolve the chemical and temperature gradients and the vertical extent of the line forming region in the disk.
The predicted FWHM and peak separation do not depend on the model resolution while the line fluxes are lower by up to $\sim30$\% when increasing the resolution.

The line profiles and fluxes are obtained by first solving the statistical equilibrium with 2D escape probability to obtain the level populations, and then using 2D ray-tracing.
As the three brightest lines of each CN 5--4 J--J' transition are blended and the three undetected components are more than one order of magnitude fainter, 
the model computes the sum of the CN 5--4 J=9/2--7/2 and J=11/2--9/2 lines.
The results obtained for the grid of models (Figure \ref{fig:grid}) show that the CN line fluxes depend mainly on the disk size and increase for increasing outer radii.
This is due to the fact that, as suggested by their profiles, the CN lines originate from the outer region of the disk. 
The best fit of the CN line fluxes is obtained for $R_{\rm out}=100-110$~AU.
However, as model-predicted fluxes are affected by $\sim30$\% uncertainty, values of $R_{\rm out}$ between $90-120$~AU produce CN line fluxes which are still in agreement with the observations. 
The ratio between the 566.9 and the 566.7 GHz lines predicted by the models, $R_{\rm mod}=1.2$, is slightly lower but still in agreement with the observed one ($R_{\rm obs}=1.4\pm0.2$). 
The peak separation and FWHM predicted by the models depend on both $R_{\rm out}$ and $i$.
For $R_{\rm out}=90-120$~AU the observed $\Delta V_{\rm sep}$ and FWHM are reproduced if the disk inclination is low ($i=20-30$\degr).
This is in agreement with the conclusions by \citet{akeson02} and \citet{ratzka09} based on the modeling of the SED and near-, mid-infrared visibilities.
Larger inclination angles of $40-45$\degr\ \citep{stapelfeldt98,guilloteau11} are excluded based on the observed CN profiles.

Following the results obtained from the grid of models we adopt $R_{\rm out}=110$~AU, $i=25$\degr\ and compute the fluxes and profiles of all the lines covered by our observations using the high resolution model. 
Figure \ref{fig:CN_model} shows the region in the disk where 50\% of the CN N=5--4 J=9/2--7/2 and J=11/2--9/2 lines originate according to our model.
This is obtained using vertical escape probability and without accounting for disk inclination.
The model indicates that the CN lines are excited in a disk upper layer located at 40--110 AU distance from the star, which is irradiated by the stellar UV field. 
This heats the gas up to temperatures of $\sim50-700$ K, while the gas density is $\sim3~10^7-3~10^9$~\cmc, thus the CN lines are almost thermalized and optically thick ($\tau\sim10^2-10^3$) in the line emitting region.
In this disk layer, CN is produced by H + CN$^{+}$, and C + NO reactions, and photo-dissociation of HCN, while the main destruction route is photo-dissociation of CN. 

In Figure \ref{fig:hifi_lines} and Table \ref{tab:lines} the model-predicted line profiles and fluxes are compared with the observed ones.
For CN lines, the model predicts a FWHM of 3.9~\kms\ and peak separation of 2.6 \kms, in agreement with observations (FWHM$=4.0,4.2\pm0.5$~\kms, $\Delta V_{\rm sep}=2.4,2.1\pm0.5$ \kms). 
Also the line fluxes agree with observations within a factor $\sim0.9,1.2$.
The sum of the fluxes of the N=2--1 components, instead, is about an order of magnitude lower than what observed by \citet{guilloteau13}, who suggest that the line is dominated by envelope emission.
The CO, $^{13}$CO, \ho, and SO line fluxes predicted by the disk model are from a factor four to two orders of magnitude smaller than observed, as well as for the high-J CO, \ho, OH, and atomic \oi, \cii\ lines previously observed with PACS  (see Table 4 by \citealt{podio12}). 
This is a further evidence that these lines are dominated by envelope and outflow emission, as already suggested by the fact that \oi, \cii\ emission is spatially extended with PACS ($\ge9$\arcsec) and by the broad profiles of CO, \ho, SO, and \cii\ obtained with HIFI. \\

\section{Conclusions}
\label{sect:conclusions}

The {\it Herschel}/HIFI observations of the T Tau system show emission in a number of molecular and atomic lines. 
The origin of the emission lines in embedded, accreting/ejecting sources is highly debated, being crucial to identify a tracer to dig out the faint disk emission \citep[e.g., ][]{podio12,podio13}.
In the case of T Tau, the CO, \ho, and \cii\ lines clearly trace high-velocity outflowing gas. 
By contrast with those lines and with previously observed CN 2--1 lines \citep{guilloteau13}, the CN 5--4 lines show narrow double-peaked profiles centered at the systemic velocity, suggesting an origin in the outer disk of T Tau N.
Disk modeling 
predicts CN line fluxes and profiles in agreement with observed ones
and constrains the size of the gaseous disk of T Tau N ($R_{\rm out}=110^{+10}_{-20}$~AU) and its inclination ($i=25$\degr$\pm5$\degr).
This study demonstrates that high-J CN lines are a unique tool to probe the gaseous disk of strongly accreting/ejecting sources 
and paves the way for future observations of embedded disks with ALMA.

\begin{acknowledgements}
LP and SDB acknowledge funding from the European FP7 (PIEF-GA-2009-253896)
and the National Science Foundation (AST-0954811).
\end{acknowledgements}


\begin{thebibliography}{49}
\expandafter\ifx\csname natexlab\endcsname\relax\def\natexlab#1{#1}\fi

\bibitem[{{Akeson} {et~al.}(2002){Akeson}, {Ciardi}, {van Belle}, \&
  {Creech-Eakman}}]{akeson02}
{Akeson}, R.~L., {Ciardi}, D.~R., {van Belle}, G.~T., \& {Creech-Eakman}, M.~J.
  2002, \apj, 566, 1124

\bibitem[{{Akeson} {et~al.}(1998){Akeson}, {Koerner}, \& {Jensen}}]{akeson98}
{Akeson}, R.~L., {Koerner}, D.~W., \& {Jensen}, E.~L.~N. 1998, \apj, 505, 358

\bibitem[{{Aresu} {et~al.}(2011){Aresu}, {Kamp}, {Meijerink}, {Woitke}, {Thi},
  \& {Spaans}}]{aresu11}
{Aresu}, G., {Kamp}, I., {Meijerink}, R., {et~al.} 2011, \aap, 526, A163

\bibitem[{{Beckwith} {et~al.}(1990){Beckwith}, {Sargent}, {Chini}, \&
  {Guesten}}]{beckwith90}
{Beckwith}, S.~V.~W., {Sargent}, A.~I., {Chini}, R.~S., \& {Guesten}, R. 1990,
  \aj, 99, 924

\bibitem[{{Bohm} \& {Solf}(1994)}]{bohm94}
{Bohm}, K.-H., \& {Solf}, J. 1994, \apj, 430, 277

\bibitem[{{Calvet} {et~al.}(2004){Calvet}, {Muzerolle}, {Brice{\~n}o},
  {Hern{\'a}ndez}, {Hartmann}, {Saucedo}, \& {Gordon}}]{calvet04}
{Calvet}, N., {Muzerolle}, J., {Brice{\~n}o}, C., {et~al.} 2004, \aj, 128, 1294

\bibitem[{{Chapillon} {et~al.}(2012){Chapillon}, {Guilloteau}, {Dutrey},
  {Pi{\'e}tu}, \& {Gu{\'e}lin}}]{chapillon12}
{Chapillon}, E., {Guilloteau}, S., {Dutrey}, A., {Pi{\'e}tu}, V., \&
  {Gu{\'e}lin}, M. 2012, \aap, 537, A60

\bibitem[{{de Graauw} {et~al.}(2010){de Graauw}, {Helmich}, {Phillips},
  {Stutzki}, {Caux}, {Whyborn}, {Dieleman}, {Roelfsema}, {Aarts}, {Assendorp},
  {Bachiller}, {Baechtold}, {Barcia}, {Beintema}, {Belitsky}, {Benz}, {Bieber},
  {Boogert}, {Borys}, {Bumble}, {Ca{\"i}s}, {Caris}, {Cerulli-Irelli},
  {Chattopadhyay}, {Cherednichenko}, {Ciechanowicz}, {Coeur-Joly}, {Comito},
  {Cros}, {de Jonge}, {de Lange}, {Delforges}, {Delorme}, {den Boggende},
  {Desbat}, {Diez-Gonz{\'a}lez}, {di Giorgio}, {Dubbeldam}, {Edwards},
  {Eggens}, {Erickson}, {Evers}, {Fich}, {Finn}, {Franke}, {Gaier}, {Gal},
  {Gao}, {Gallego}, {Gauffre}, {Gill}, {Glenz}, {Golstein}, {Goulooze},
  {Gunsing}, {G{\"u}sten}, {Hartogh}, {Hatch}, {Higgins}, {Honingh}, {Huisman},
  {Jackson}, {Jacobs}, {Jacobs}, {Jarchow}, {Javadi}, {Jellema}, {Justen},
  {Karpov}, {Kasemann}, {Kawamura}, {Keizer}, {Kester}, {Klapwijk}, {Klein},
  {Kollberg}, {Kooi}, {Kooiman}, {Kopf}, {Krause}, {Krieg}, {Kramer},
  {Kruizenga}, {Kuhn}, {Laauwen}, {Lai}, {Larsson}, {Leduc}, {Leinz}, {Lin},
  {Liseau}, {Liu}, {Loose}, {L{\'o}pez-Fernandez}, {Lord}, {Luinge}, {Marston},
  {Mart{\'{\i}}n-Pintado}, {Maestrini}, {Maiwald}, {McCoey}, {Mehdi}, {Megej},
  {Melchior}, {Meinsma}, {Merkel}, {Michalska}, {Monstein}, {Moratschke},
  {Morris}, {Muller}, {Murphy}, {Naber}, {Natale}, {Nowosielski}, {Nuzzolo},
  {Olberg}, {Olbrich}, {Orfei}, {Orleanski}, {Ossenkopf}, {Peacock}, {Pearson},
  {Peron}, {Phillip-May}, {Piazzo}, {Planesas}, {Rataj}, {Ravera}, {Risacher},
  {Salez}, {Samoska}, {Saraceno}, {Schieder}, {Schlecht}, {Schl{\"o}der},
  {Schm{\"u}lling}, {Schultz}, {Schuster}, {Siebertz}, {Smit}, {Szczerba},
  {Shipman}, {Steinmetz}, {Stern}, {Stokroos}, {Teipen}, {Teyssier}, {Tils},
  {Trappe}, {van Baaren}, {van Leeuwen}, {van de Stadt}, {Visser}, {Wildeman},
  {Wafelbakker}, {Ward}, {Wesselius}, {Wild}, {Wulff}, {Wunsch}, {Tielens},
  {Zaal}, {Zirath}, {Zmuidzinas}, \& {Zwart}}]{degraauw10}
{de Graauw}, T., {Helmich}, F.~P., {Phillips}, T.~G., {et~al.} 2010, \aap, 518,
  L6

\bibitem[{{Draine} \& {Lee}(1984)}]{draine84}
{Draine}, B.~T., \& {Lee}, H.~M. 1984, \apj, 285, 89

\bibitem[{{Duch{\^e}ne} {et~al.}(2005){Duch{\^e}ne}, {Ghez}, {McCabe}, \&
  {Ceccarelli}}]{duchene05}
{Duch{\^e}ne}, G., {Ghez}, A.~M., {McCabe}, C., \& {Ceccarelli}, C. 2005, \apj,
  628, 832

\bibitem[{{Dutrey} {et~al.}(1997){Dutrey}, {Guilloteau}, \&
  {Guelin}}]{dutrey97}
{Dutrey}, A., {Guilloteau}, S., \& {Guelin}, M. 1997, \aap, 317, L55

\bibitem[{{Dutrey} {et~al.}(1998){Dutrey}, {Guilloteau}, {Prato}, {Simon},
  {Duvert}, {Schuster}, \& {Menard}}]{dutrey98}
{Dutrey}, A., {Guilloteau}, S., {Prato}, L., {et~al.} 1998, \aap, 338, L63

\bibitem[{{Edwards} \& {Snell}(1982)}]{edwards82}
{Edwards}, S., \& {Snell}, R.~L. 1982, \apj, 261, 151

\bibitem[{{Eisl{\"o}ffel} \& {Mundt}(1998)}]{eisloffel98}
{Eisl{\"o}ffel}, J., \& {Mundt}, R. 1998, \aj, 115, 1554

\bibitem[{{Furlan} {et~al.}(2006){Furlan}, {Hartmann}, {Calvet}, {D'Alessio},
  {Franco-Hern{\'a}ndez}, {Forrest}, {Watson}, {Uchida}, {Sargent}, {Green},
  {Keller}, \& {Herter}}]{furlan06}
{Furlan}, E., {Hartmann}, L., {Calvet}, N., {et~al.} 2006, \apjs, 165, 568

\bibitem[{{Gorti} {et~al.}(2011){Gorti}, {Hollenbach}, {Najita}, \&
  {Pascucci}}]{gorti11}
{Gorti}, U., {Hollenbach}, D., {Najita}, J., \& {Pascucci}, I. 2011, \apj, 735,
  90

\bibitem[{{G{\"u}del} {et~al.}(2007){G{\"u}del}, {Telleschi}, {Audard},
  {Skinner}, {Briggs}, {Palla}, \& {Dougados}}]{gudel07a}
{G{\"u}del}, M., {Telleschi}, A., {Audard}, M., {et~al.} 2007, \aap, 468, 515

\bibitem[{{Guilloteau} {et~al.}(2013){Guilloteau}, {Di Folco}, {Dutrey},
  {Simon}, {Grosso}, \& {Pi{\'e}tu}}]{guilloteau13}
{Guilloteau}, S., {Di Folco}, E., {Dutrey}, A., {et~al.} 2013, \aap, 549, A92

\bibitem[{{Guilloteau} {et~al.}(2011){Guilloteau}, {Dutrey}, {Pi{\'e}tu}, \&
  {Boehler}}]{guilloteau11}
{Guilloteau}, S., {Dutrey}, A., {Pi{\'e}tu}, V., \& {Boehler}, Y. 2011, \aap,
  529, A105

\bibitem[{{Gustafsson} {et~al.}(2008){Gustafsson}, {Labadie}, {Herbst}, \&
  {Kasper}}]{gustafsson08}
{Gustafsson}, M., {Labadie}, L., {Herbst}, T.~M., \& {Kasper}, M. 2008, \aap,
  488, 235

\bibitem[{{Harris} {et~al.}(2012){Harris}, {Andrews}, {Wilner}, \&
  {Kraus}}]{harris12}
{Harris}, R.~J., {Andrews}, S.~M., {Wilner}, D.~J., \& {Kraus}, A.~L. 2012,
  \apj, 751, 115

\bibitem[{{Herbst} {et~al.}(2007){Herbst}, {Hartung}, {Kasper}, {Leinert}, \&
  {Ratzka}}]{herbst07}
{Herbst}, T.~M., {Hartung}, M., {Kasper}, M.~E., {Leinert}, C., \& {Ratzka}, T.
  2007, \aj, 134, 359

\bibitem[{{Hogerheijde} {et~al.}(1997){Hogerheijde}, {van Langevelde}, {Mundy},
  {Blake}, \& {van Dishoeck}}]{hogerheijde97}
{Hogerheijde}, M.~R., {van Langevelde}, H.~J., {Mundy}, L.~G., {Blake}, G.~A.,
  \& {van Dishoeck}, E.~F. 1997, \apjl, 490, L99

\bibitem[{{Kamp} {et~al.}(2010){Kamp}, {Tilling}, {Woitke}, {Thi}, \&
  {Hogerheijde}}]{kamp10}
{Kamp}, I., {Tilling}, I., {Woitke}, P., {Thi}, W.-F., \& {Hogerheijde}, M.
  2010, \aap, 510, A18

\bibitem[{{K{\"o}hler} {et~al.}(2008){K{\"o}hler}, {Ratzka}, {Herbst}, \&
  {Kasper}}]{kohler08}
{K{\"o}hler}, R., {Ratzka}, T., {Herbst}, T.~M., \& {Kasper}, M. 2008, \aap,
  482, 929

\bibitem[{{Kristensen} {et~al.}(2012){Kristensen}, {van Dishoeck}, {Bergin},
  {Visser}, {Y{\i}ld{\i}z}, {San Jose-Garcia}, {J{\o}rgensen}, {Herczeg},
  {Johnstone}, {Wampfler}, {Benz}, {Bruderer}, {Cabrit}, {Caselli}, {Doty},
  {Harsono}, {Herpin}, {Hogerheijde}, {Karska}, {van Kempen}, {Liseau},
  {Nisini}, {Tafalla}, {van der Tak}, \& {Wyrowski}}]{kristensen12}
{Kristensen}, L.~E., {van Dishoeck}, E.~F., {Bergin}, E.~A., {et~al.} 2012,
  \aap, 542, A8

\bibitem[{{Meijerink} {et~al.}(2012){Meijerink}, {Aresu}, {Kamp}, {Spaans},
  {Thi}, \& {Woitke}}]{meijerink12}
{Meijerink}, R., {Aresu}, G., {Kamp}, I., {et~al.} 2012, \aap, 547, A68

\bibitem[{{Milam} {et~al.}(2005){Milam}, {Savage}, {Brewster}, {Ziurys}, \&
  {Wyckoff}}]{milam05}
{Milam}, S.~N., {Savage}, C., {Brewster}, M.~A., {Ziurys}, L.~M., \& {Wyckoff},
  S. 2005, \apj, 634, 1126

\bibitem[{{M{\"u}ller} {et~al.}(2001){M{\"u}ller}, {Thorwirth}, {Roth}, \&
  {Winnewisser}}]{muller01}
{M{\"u}ller}, H.~S.~P., {Thorwirth}, S., {Roth}, D.~A., \& {Winnewisser}, G.
  2001, \aap, 370, L49

\bibitem[{{{\"O}berg} {et~al.}(2010){{\"O}berg}, {Qi}, {Fogel}, {Bergin},
  {Andrews}, {Espaillat}, {van Kempen}, {Wilner}, \& {Pascucci}}]{oberg10}
{{\"O}berg}, K.~I., {Qi}, C., {Fogel}, J.~K.~J., {et~al.} 2010, \apj, 720, 480

\bibitem[{{{\"O}berg} {et~al.}(2011){{\"O}berg}, {Qi}, {Fogel}, {Bergin},
  {Andrews}, {Espaillat}, {Wilner}, {Pascucci}, \& {Kastner}}]{oberg11}
{{\"O}berg}, K.~I., {Qi}, C., {Fogel}, J.~K.~J., {et~al.} 2011, \apj, 734, 98

\bibitem[{{Pickett} {et~al.}(1998){Pickett}, {Poynter}, {Cohen}, {Delitsky},
  {Pearson}, \& {M{\"u}ller}}]{pickett98}
{Pickett}, H.~M., {Poynter}, R.~L., {Cohen}, E.~A., {et~al.} 1998, \jqsrt, 60,
  883

\bibitem[{{Pi{\'e}tu} {et~al.}(2007){Pi{\'e}tu}, {Dutrey}, \&
  {Guilloteau}}]{pietu07}
{Pi{\'e}tu}, V., {Dutrey}, A., \& {Guilloteau}, S. 2007, \aap, 467, 163

\bibitem[{{Pilbratt} {et~al.}(2010){Pilbratt}, {Riedinger}, {Passvogel},
  {Crone}, {Doyle}, {Gageur}, {Heras}, {Jewell}, {Metcalfe}, {Ott}, \&
  {Schmidt}}]{pilbratt10}
{Pilbratt}, G.~L., {Riedinger}, J.~R., {Passvogel}, T., {et~al.} 2010, \aap,
  518, L1

\bibitem[{{Podio} {et~al.}(2012){Podio}, {Kamp}, {Flower}, {Howard}, {Sandell},
  {Mora}, {Aresu}, {Brittain}, {Dent}, {Pinte}, \& {White}}]{podio12}
{Podio}, L., {Kamp}, I., {Flower}, D., {et~al.} 2012, \aap, 545, A44

\bibitem[{{Podio} {et~al.}(2013){Podio}, {Kamp}, {Codella}, {Cabrit}, {Nisini},
  {Dougados}, {Sandell}, {Williams}, {Testi}, {Thi}, {Woitke}, {Meijerink},
  {Spaans}, {Aresu}, {M{\'e}nard}, \& {Pinte}}]{podio13}
{Podio}, L., {Kamp}, I., {Codella}, C., {et~al.} 2013, \apjl, 766, L5

\bibitem[{{Qi} {et~al.}(2004){Qi}, {Ho}, {Wilner}, {Takakuwa}, {Hirano},
  {Ohashi}, {Bourke}, {Zhang}, {Blake}, {Hogerheijde}, {Saito}, {Choi}, \&
  {Yang}}]{qi04}
{Qi}, C., {Ho}, P.~T.~P., {Wilner}, D.~J., {et~al.} 2004, \apjl, 616, L11

\bibitem[{{Ratzka} {et~al.}(2009){Ratzka}, {Schegerer}, {Leinert},
  {{\'A}brah{\'a}m}, {Henning}, {Herbst}, {K{\"o}hler}, {Wolf}, \&
  {Zinnecker}}]{ratzka09}
{Ratzka}, T., {Schegerer}, A.~A., {Leinert}, C., {et~al.} 2009, \aap, 502, 623

\bibitem[{{Roelfsema} {et~al.}(2012){Roelfsema}, {Helmich}, {Teyssier},
  {Ossenkopf}, {Morris}, {Olberg}, {Shipman}, {Risacher}, {Akyilmaz},
  {Assendorp}, {Avruch}, {Beintema}, {Biver}, {Boogert}, {Borys}, {Braine},
  {Caris}, {Caux}, {Cernicharo}, {Coeur-Joly}, {Comito}, {de Lange},
  {Delforge}, {Dieleman}, {Dubbeldam}, {de Graauw}, {Edwards}, {Fich},
  {Flederus}, {Gal}, {di Giorgio}, {Herpin}, {Higgins}, {Hoac}, {Huisman},
  {Jarchow}, {Jellema}, {de Jonge}, {Kester}, {Klein}, {Kooi}, {Kramer},
  {Laauwen}, {Larsson}, {Leinz}, {Lord}, {Lorenzani}, {Luinge}, {Marston},
  {Mart{\'{\i}}n-Pintado}, {McCoey}, {Melchior}, {Michalska}, {Moreno},
  {M{\"u}ller}, {Nowosielski}, {Okada}, {Orlea{\'n}ski}, {Phillips}, {Pearson},
  {Rabois}, {Ravera}, {Rector}, {Rengel}, {Sagawa}, {Salomons},
  {S{\'a}nchez-Su{\'a}rez}, {Schieder}, {Schl{\"o}der}, {Schm{\"u}lling},
  {Soldati}, {Stutzki}, {Thomas}, {Tielens}, {Vastel}, {Wildeman}, {Xie},
  {Xilouris}, {Wafelbakker}, {Whyborn}, {Zaal}, {Bell}, {Bjerkeli}, {De Beck},
  {Cavali{\'e}}, {Crockett}, {Hily-Blant}, {Kama}, {Kaminski}, {Lefl{\'o}ch},
  {Lombaert}, {de Luca}, {Makai}, {Marseille}, {Nagy}, {Pacheco}, {van der
  Wiel}, {Wang}, \& {Y{\i}ld{\i}z}}]{roelfsema12}
{Roelfsema}, P.~R., {Helmich}, F.~P., {Teyssier}, D., {et~al.} 2012, \aap, 537,
  A17

\bibitem[{{Schuster} {et~al.}(1993){Schuster}, {Harris}, {Anderson}, \&
  {Russell}}]{schuster93}
{Schuster}, K.~F., {Harris}, A.~I., {Anderson}, N., \& {Russell}, A.~P.~G.
  1993, \apjl, 412, L67

\bibitem[{{Solf} \& {B{\"o}hm}(1999)}]{solf99}
{Solf}, J., \& {B{\"o}hm}, K.-H. 1999, \apj, 523, 709

\bibitem[{{Spinoglio} {et~al.}(2000){Spinoglio}, {Giannini}, {Nisini}, {van den
  Ancker}, {Caux}, {Di Giorgio}, {Lorenzetti}, {Palla}, {Pezzuto}, {Saraceno},
  {Smith}, \& {White}}]{spinoglio00}
{Spinoglio}, L., {Giannini}, T., {Nisini}, B., {et~al.} 2000, \aap, 353, 1055

\bibitem[{{Stapelfeldt} {et~al.}(1998){Stapelfeldt}, {Burrows}, {Krist},
  {Watson}, {Ballester}, {Clarke}, {Crisp}, {Evans}, {Gallagher}, {Griffiths},
  {Hester}, {Hoessel}, {Holtzman}, {Mould}, {Scowen}, {Trauger}, \&
  {Westphal}}]{stapelfeldt98}
{Stapelfeldt}, K.~R., {Burrows}, C.~J., {Krist}, J.~E., {et~al.} 1998, \apj,
  508, 736

\bibitem[{{Thi} {et~al.}(2004){Thi}, {van Zadelhoff}, \& {van
  Dishoeck}}]{thi04}
{Thi}, W.-F., {van Zadelhoff}, G.-J., \& {van Dishoeck}, E.~F. 2004, \aap, 425,
  955

\bibitem[{{Thi} {et~al.}(2010){Thi}, {Mathews}, {M{\'e}nard}, {Woitke},
  {Meeus}, {Riviere-Marichalar}, {Pinte}, {Howard}, {Roberge}, {Sandell},
  {Pascucci}, {Riaz}, {Grady}, {Dent}, {Kamp}, {Duch{\^e}ne}, {Augereau},
  {Pantin}, {Vandenbussche}, {Tilling}, {Williams}, {Eiroa}, {Barrado},
  {Alacid}, {Andrews}, {Ardila}, {Aresu}, {Brittain}, {Ciardi}, {Danchi},
  {Fedele}, {de Gregorio-Monsalvo}, {Heras}, {Huelamo}, {Krivov}, {Lebreton},
  {Liseau}, {Martin-Zaidi}, {Mendigut{\'{\i}}a}, {Montesinos}, {Mora},
  {Morales-Calderon}, {Nomura}, {Phillips}, {Podio}, {Poelman}, {Ramsay},
  {Rice}, {Solano}, {Walker}, {White}, \& {Wright}}]{thi10}
{Thi}, W.-F., {Mathews}, G., {M{\'e}nard}, F., {et~al.} 2010, \aap, 518, L125

\bibitem[{{van Boekel} {et~al.}(2009){van Boekel}, {G{\"u}del}, {Henning},
  {Lahuis}, \& {Pantin}}]{vanboekel09}
{van Boekel}, R., {G{\"u}del}, M., {Henning}, T., {Lahuis}, F., \& {Pantin}, E.
  2009, \aap, 497, 137

\bibitem[{{White} \& {Ghez}(2001)}]{white01}
{White}, R.~J., \& {Ghez}, A.~M. 2001, \apj, 556, 265

\bibitem[{{Woitke} {et~al.}(2009){Woitke}, {Kamp}, \& {Thi}}]{woitke09}
{Woitke}, P., {Kamp}, I., \& {Thi}, W.-F. 2009, \aap, 501, 383

\bibitem[{{Zubko} {et~al.}(1996){Zubko}, {Mennella}, {Colangeli}, \&
  {Bussoletti}}]{zubko96}
{Zubko}, V.~G., {Mennella}, V., {Colangeli}, L., \& {Bussoletti}, E. 1996,
  \mnras, 282, 1321

\end{thebibliography}

\end{document}